\documentclass[twocolumn,amsmath,amssymb,aps,prl,floatfix,nofootinbib,superscriptaddress]{revtex4}
\usepackage{graphicx,graphics,times}
\usepackage{bm}
\usepackage{longtable}

\def\be{\begin{equation}}
\def\ee{\end{equation}}
\def\ba{\begin{eqnarray}}
\def\ea{\end{eqnarray}}
\def\la{\langle}
\def\ra{\rangle}

\begin{document}
\title{Entanglement Transfer Through an Antiferromagnetic Spin Chain}
\author{Abolfazl Bayat}
\author{Sougato Bose}
\affiliation{Department of Physics and Astronomy, University College
London, Gower St., London WC1E 6BT, UK}

\begin{abstract}
We study the possibility of using an uniformly coupled finite
antiferromagnetic spin-1/2 Heisenberg chain as a channel for
transmitting entanglement. One member of a pair of maximally
entangled spins is initially appended to one end of a chain in its
ground state and the dynamical propagation of this entanglement to
the other end is calculated. We show that compared to the analogous
scheme with a ferromagnetic chain in its ground state, here the
entanglement is transmitted faster, with less decay, with a much
higher purity and as a narrow pulse form rising non-analytically
from zero. Here non-zero temperatures and depolarizing environments
are both found to be less destructive in comparison to the
ferromagnetic case. The entanglement is found to propagate through
the chain in a peculiar fashion whereby it hops to skip alternate
sites.
\end{abstract}

\date{\today}
 \maketitle

Identifying potential methods for linking distinct quantum
processors or registers is a crucial part of scalable quantum
computing technology. Studying the potential of spin chains as
quantum wires for the above purpose has recently emerged as an area
of significant activity \cite{bose,rest,rest1,bayat,decoherence} as
they can successfully transfer quantum states and Entanglement
over short distance scales. One motivation for such wires is to
circumvent the necessity of inter-conversion between solid state
qubits and photons when connecting solid state quantum registers
separated by short distances. Additionally, spin chains are systems
of permanently coupled spins (essentially, a one dimensional magnet
or isomorphic system). Thereby studying their potential to transfer
quantum information automatically answers the question as to how
well one can accomplish the transfer of a quantum state through a
chain of coupled qubits without requiring the switchability or
tunability of any of the interactions inside the chain -- an example
of {\em minimal control} in quantum information processing. This
line of research can also be motivated simply as the study of
canonical condensed mater systems from a quantum information
perspective. As opposed to the hugely popular field examining how
much entanglement exists inside such systems \cite{vlatko}, this work
investigates how quantum information {\em passes through} such
systems.

 In the original algorithm \cite{bose}, as well as in most subsequent work
\cite{rest,rest1,bayat,decoherence}, a chain of qubits (spin-1/2
systems) initialized in a fully polarized (symmetry broken) state
plays the role of the channel. This would be the ground state, for
example, if a ferromagnetic (FM) spin chain was used as the channel.
The important noise factors such as the effects of temperature
\cite{bayat} and decoherence \cite{decoherence} have also been
investigated for such FM channels. By now a plethora of physical
implementations of such a scheme has either been performed using NMR
\cite{NMR} or suggested \cite{suggested} (for Josephson junction
arrays, trapped electron chains etc). However, how about using an
antiferromagnetic (AFM) spin chain initialized in its ground state
as a quantum channel for the transfer of entanglement? Strangely enough, the
simplest version of this, namely an uniformly coupled spin-1/2
Heisenberg AFM chain as a channel for quantum information transfer,
remains unstudied though examples of such chains are much more
common than FM chains in condensed matter, including ones on which
NMR studies are done \cite{AFM-Polymer}. They can be simulated in
optical lattices \cite{duan} and with Phosphorous doped Silicon
\cite{kane}. Most strikingly, thanks to the progress of
nanotechnology, antiferromagnetic (AFM) spin chains up to 10 spins
in length have been built experimentally recently and the spin of
the atoms and also the couplings between them can be probed
individually by scanning tunneling microscopes \cite{hirjib}. This
truly motivates an examination of the transfer of entanglement through AFM
spin chains. Additionally, compared to FM channels, one can expect
several qualitatively different features in AFM spin chain channels
as they already have lot of entanglement inside, and the monogamous nature of
shared entanglement may lead to nontrivial dynamics. Also, the channel is
rotationally fully symmetric, and this leads to a qualitatively
different channel for the transfer of quantum information.

 Recently, the quality of state and entanglement transfer through all phases of a
spin-1 chain (both FM and AFM) has been studied and some AFM phases
have been shown to outperform the FM phases as a quantum wire
\cite{sanpera}. Dimerized AFM states of Spin-1 chains can also
enable certain state transfer schemes involving an adiabatic
modulation of couplings \cite{sanpera2}. It has also been shown that
quantum information can be efficiently transferred between weakly
coupled end spins of an AFM chain because of an effective direct
coupling between these spins \cite{lorenzo}. Some other recent
studies of quantum state and entanglement transfer \cite{dimer,plenio1} and entanglement
dynamics \cite{fazio} have considered initial states deviating from
the usual fully polarized state. However, what about the simplest
AFM chain of uniformly coupled spin-1/2 systems? In this letter, we
obtain curious results about the propagation of entanglement through such a
chain, in particular that it {\em hops to skip alternate sites} and
that the entanglement transmitted through the channel rises from zero sharply
and {\em non-analytically as a narrow pulse}. Such striking features
will be very interesting to test with the finite AFM chains. In
addition, we find that a channel with AFM initial state consistently
outperforms the corresponding FM case for comparable chain lengths
and reasonable times, even when temperature and decoherence effects
are included.

  We follow the approach of \cite{bose} to transfer entanglement from
one end of an open AFM spin chain to the other and compare the
quality and behavior in different situations with the case of FM.
The Hamiltonian of the open chain with length $N_{ch}$ is
\begin{equation}\label{Hamiltonian_ch}
    H_{ch}=J\sum_{i=1}^{N_{ch}-1}\sigma_i.\sigma_{i+1}.
\end{equation}
where the $\sigma_k=(\sigma_k^x,\sigma_k^y,\sigma_k^z)$ is a
vector contains Pauli matrices which act on the site $k$ and $J$
is the coupling constant ($J>0$ for AFM and $J<0$ for FM). The
protocol for transferring the entanglement is as follows: We place a pair of
spins $0'$ and $0$ in the singlet state
$|\psi^-\ra_{0'0}=\frac{1}{\sqrt{2}}(|0\ra_{0'}|1\ra_0-|1\ra_{0'}|0\ra_{0})$
while the channel (spins 1 to $N_{ch}$) is in its ground state
$|\psi_g\ra_{ch}$ ({\em i.e.} the ground state of $H_{ch}$). Note
that for the AFM case, $|\psi_g\ra_{ch}$ is a global singlet state
of $N_{ch}$ spins, while for the FM case it is a fully polarized
ground state with all spins pointing in a given direction. Also
note that both for the AFM chain for odd $N_{ch}$, and the FM
chain, the ground state is non-unique and a unique ground state
$|\psi_g\ra_{ch}$ is selected out by applying an arbitrarily small
magnetic field which does not affect the eigenvectors and just split up the degenerate energies. If one avoids applying the magnetic field to choose a unique ground state, any superposition
of the degenerate eigenvectors could be chosen for the initial state. So that we can not get a unique result for comparison to other chains and also
since in this situation the mixedness of the final state is increased the quality of
entanglement goes down. When the initial state is prepared, we then turn on the interaction between spin $0$
and first spin of the channel (spin $1$). The Hamiltonian
including this additional interaction is
\begin{equation}\label{Hamiltonain}
    H=I_{0'}\otimes(J\sigma_0.\sigma_1+H_{ch}).
\end{equation}
The total length of the system considered is thus $N=N_{ch}+2$ with
the total
Hamiltonian being $H$ (so that $0'$ never interacts with the
channel) and the initial state being
\begin{equation}\label{init}
   |\psi(0)\ra=|\psi^-\ra_{0'0}\otimes|\psi_g\ra_{ch}.
\end{equation}
 We are interested at the times that the
 entanglement between the spins $0'$ and $N_{ch}$ peaks, which is the aim of the entanglement distribution through our spin chain channel.  By turning on
the interaction between spin $0$ and the first spin of the channel
(spin 1) the initial state evolves to the state
$|\psi(t)\ra=e^{-iHt}|\psi(0)\ra$ and one can compute the density
matrix $\rho_{0'N_{ch}}=tr_{\hat{0'N_{ch}}}\{|\psi(t)\ra\la
\psi(t)|\}$ where the meaning of $tr_{\hat{ij}}$ is the trace over
whole of the system {\em except} sites $i$ and $j$.
\begin{figure}
\centering
    \includegraphics[width=7cm,height=6cm,angle=0]{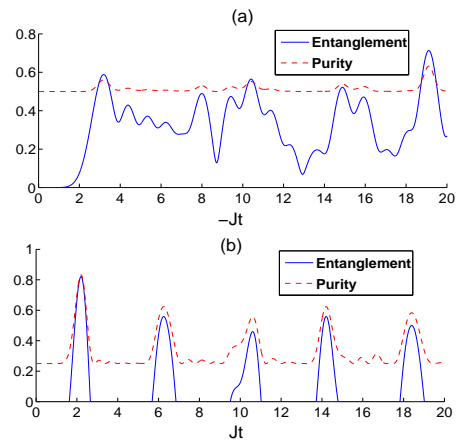}
    \caption{(Color online) The concurrence and the purity of the state $\rho_{0'N_{ch}}$ for the chain of length 10. Figure (a) is for the case of
     FM chain ($J<0$) and figure (b) is for AFM one ($J>0$). }
     \label{fig1}
\end{figure}

 In figure \ref{fig1} the entanglement (as quantified by the entanglement
concurrence \cite{concurrence}) and the purity (as quantified by
$tr(\rho_{0'N_{ch}}^2)$ of the state $\rho_{0'N_{ch}}$ for both
the cases of AFM and FM chain as a function of time have been
plotted for a system of length $N=10$. It is clear from the
figures that the behavior of the entanglement and the purity of the
entangled state is completely different for the two cases. For the
much studied FM case \cite{bose}, the entanglement of the spins $0'$ and
$N_{ch}$ is simply equal to the modulus of the amplitude of an
excitation to transfer from the site $0$ to the site $N_{ch}$ due
to $H$ which is always an analytic function. In contrast, in the
AFM case we find a {\em nonanalytic} behavior in entanglement as a function
of time. It is zero for most of the time and at regular intervals
it suddenly grows up and makes a peak with its derivative being
discontinuous at the point it starts to rise from zero. This
behavior can be understood by realizing that the channel has to
act as a purely depolarizing channel (equally probable random
actions of all the three Pauli operators on a state while it
passes through the channel) because of the $SU(2)$ symmetry of the
channel state $|\psi_g\ra_{ch}$. When one member of an entangled
pair of qubits is transmitted through such a channel, then the two
qubit state evolves to a Werner state \cite{BDSW}
\begin{equation}\label{werner}
   \rho_{0'N_{ch}}(t)=p(t)|\psi^-\ra\la\psi^-|+(1-p(t))I/4,
\end{equation}
where $I$ is the identity
matrix for two qubits, and $p(t)$ is a time-dependent positive number $\leq 1$
parameterizing the state. $\rho_{0'N_{ch}}$ will thus always be a
Werner state with its parameter $p$ varying with time. Initially
$p$ is zero (both qubits $0'$ and $N_{ch}$ are maximally entangled
to distinct systems $0'$ with $0$ and $N_{ch}$ with the rest of
the chain respectively) and it rises from that as a simple
trigonometric function of time. For example, for the simplest case
$N_{ch}=2$ (for which $|\psi_g\ra$ is trivially a singlet and the
starting state for the whole four qubit system is
$|\psi(0)\ra=|\psi^-\ra \otimes |\psi^-\ra$), one can analytically
calculate the evolution easily to obtain $p(t)=\sin^22Jt$. It is
known that as long as $p$ remains $\leq 1/3$, the entanglement of the final state (\ref{werner})
stays constant at zero \cite{BDSW}, and thereby the curve for entanglement
versus time has a vanishing derivative. The entanglement starts to rise
suddenly as soon as $p$ exceeds $1/3$, but the trigonometric form
of $p$ (such as $\sin^22Jt$ for $N_{ch}=2$) does not have a
vanishing derivative (i.e., be in a maximum or minimum) at this
point. There is thus a sudden discontinuity in the derivative of
the curve of the entanglement of $0'$ and $N_{ch}$ versus time.
Though, finding a non-analyticity in the entanglement is not very interesting since
the concurrence, similar to any other entanglement measure, has the source of non-analyticity in its definition but it is worthwhile to
point out that entanglement gained by the FM chain is always analytic because in this case the correlation functions are bounded from below.

 Another important difference between the case of FM and AFM chain
that can be seen in figure \ref{fig1} is the purity of the final
entangled state. The purity of the state $\rho_{0'N_{ch}}$ is higher
in the case of AFM chain in comparison with FM one. Having a purer
entangled state transmitted is a distinct advantage as in the end
one needs a purify the transmitted states by local actions to obtain
a smaller number of states arbitrarily close to a singlet through a
process called entanglement distillation \cite{BDSW}. Only such purified entanglement is
really useful for linking distinct quantum processors and purer the
shared entangled state, less is the effort to distill it. In
addition, the very fact that $\rho_{0'N_{ch}}$ is a Werner state is
a distinct advantage compared to the FM case. Werner states are a
class of mixed states for which entanglement distillation methods are very
well developed right from the start to the extent that in the
original entanglement distillation paper \cite{BDSW} it was proposed to
convert any mixed state to a Werner state first and then distill pure entanglement from it.

For the AFM chain with even number of spins since the final state is always a Werner 
state with the form of (\ref{werner}), entanglement and purity are
uniquely determined by the parameter $p$. It is easy to show that the concurrence 
of the state (\ref{werner}) is $(3p-1)/2$ and its purity is $(3p^2+1)/4$. 
For quantum state transferring and quantum communication one might prefer to
directly send quantum states through the chain \cite{bose}. 
In this case one generates an arbitrary state $|\psi_s\ra=\alpha|0\ra+\beta |1\ra$
at spin 0 while the chain (i.e., spins 1,2,...,$N_{ch}$) is in its ground state. Like the 
strategy explained above for entanglement distribution, at $t=0$ the interaction 
between spin 0 and the rest is switched on. The dynamics of the system transfers the state
$|\psi_s\ra$ through the chain till it reaches the end. So then, at some proper times state of the last site $\rho_{_{N_{ch}}}(t)$ 
is similar to $|\psi_s\ra$. One can easily compute the fidelity $F=\la\psi_s|\rho_{_{N_{ch}}}(t)|\psi_s\ra$
which is obviously a function of $\alpha$, $\beta$ and time $t$. 
One can average the fidelity $F$ over all possible input states. This can be done by averaging over the surface of the Bloch sphere
to get an input state independent parameter $F_{av}$.
An straight forward computation gives $F_{av}=(3p+1)/4$ for even AFM chains which 
is again determined uniquely by the Werner parameter $p(t)$ in Eq. (\ref{werner}). 
Thus these quantities, i.e., entanglement, purity and average fidelity, are not really independent and considering
one of them provides enough information for the others so that we mainly focus on the entanglement in this paper.

To understand the difference between AFM and FM chains it is very important to notice that when the sign of the coupling $J$
is changed the eigenvectors of the Hamiltonian do not change. So only the eigenvalues vary and consequently the ground state of the system changes.
Our investigation shows that what is really important in the dynamics
is the eigenvector which is chosen as the initial state and the sign of $J$ is not important.
It means that if for a AFM chain, which $J$ is positive, we prepare a FM initial eigenvector, which all spins are aligned into a same direction,
then the results are similar to a FM chain even though the Hamiltonian is AFM. Same results hold for the case that
we generate the AFM eigenvector as the initial state of a FM chain.
Using the AFM (FM) Hamiltonian for generating an AFM (FM) ground state have this benefit that simply with cooling
the system it goes to its ground state while for an AFM (FM) chain generating a FM (AFM) eigenvector is practically very hard and needs
lots of external control.

\begin{figure}
\centering
    \includegraphics[width=9cm,height=7cm,angle=0]{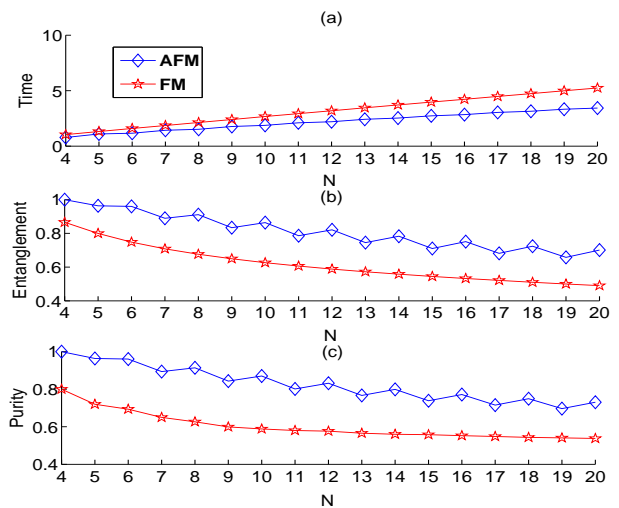}
    \caption{(Color online)In this figure we have plotted the time
    of first maximum
    and the amount of entanglement and also purity versus length
    for both cases FM and AFM chains versus the length of the chain.}
    \label{fig2}
\end{figure}
In practice, the time which one can afford to wait for the entanglement
between $0'$ and $N_{ch}$ to attain a peak is restricted by
practical considerations such as the decoherence time of the
system and simply by how much delay we can afford while connecting
quantum processors. So we restrict ourselves to the case of the
first maximum of the entanglement in time. In figure \ref{fig2}.a, we have
plotted the time that entanglement achieves its first maximum value versus
the total length of the system for chain lengths of up to $N=20$
spins for both AFM and FM chains. It is clear that the speed of entanglement
transmission through the AFM chain is higher than that through a
FM chain independent of the length of the chain. In figure
\ref{fig2}.b and \ref{fig2}.c, the amount of entanglement and purity in the
first maximum of entanglement has been compared for both of the AFM and FM
case, from which it is clear the the entanglement transmitted in the case of
AFM chain has a higher value and also it's more pure than the entanglement
transmitted in the case of FM chain. To see clearly the reason for
the above superiority of the AFM chain over the FM chain, it is
instructive to define something like a signal propagation wave in
the two cases. This is because, in the end, it is the transfer of
the state of spin $0$ to spin $N_{ch}$ that causes the entanglement to be
set up between $0'$ and $N_{ch}$. In the case of the FM chain it
is easy to define this as simply the propagation of a localized
spin flip excitation (a superposition of all one magnon states)
over a polarized background state \cite{bose}. In the case of AFM
chain it can be defined as a wave of modulation of the local
density matrix of the spins if any state is appended to one end of
the chain. For example, in an AFM ground state, the local density
matrices of each spin will be the identity matrix. However, if a
state $|1\rangle$ is appended to one end of it, and the system is
allowed to evolve in time, there will be a wave of deviation of
the local density matrices from the identity towards
$|1\rangle\langle 1|$ which will propagate through the chain. This
wave (for the AFM) simply travels faster through the chain than
the spin flip excitation of a FM, and is responsible for the
results of Fig. \ref{fig2}a. Additionally, this wave has a
significantly lower dispersion than the corresponding case for the
FM chain, which is responsible for the higher purity and higher entanglement
for the AFM case.

\begin{figure}
\centering
    \includegraphics[width=7cm,height=5cm,angle=0]{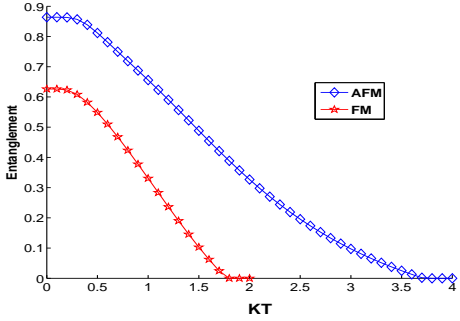}
    \caption{(Color online) The amount of the first maximum of entanglement between two ends in a chain of length 10 versus the temperature
    for both the case of FM chain and AFM one.}
    \label{fig3}
\end{figure}
 Generally when the system is in non zero temperature, the state of
the channel before evolution is described by a thermal state
$\frac{e^{-\beta H_{ch}}}{Z}$ instead of the ground state, where
$\beta=1/KT$ and $Z$ is the partition function of the channel. So in
this case the initial state of the system is
\begin{equation}\label{init_thermal}
    \rho(0)=|\psi^-\ra \la\psi^-|\otimes \frac{e^{-\beta H_{ch}}}{Z}
\end{equation}
and after time $t$
the target state $\rho_{0'N_{ch}}(t)$ can be gained by
\begin{equation}\label{rho0nt_thermal}
    \rho_{0'N_{ch}}(t)=tr_{\hat{0'N_{ch}}}\{e^{-iHt}\rho(0)e^{iHt}\}.
\end{equation}
In figure (\ref{fig3}) we have plotted the value of the first
maximum of concurrence of the state (\ref{rho0nt_thermal}) for
both the cases of FM and AFM chains in a system of length $N=10$.
The entanglement in the FM chains is more sensitive to the temperature and
decays faster than AFM chain by increasing the temperature. The
time at which the entanglement gets its first maximum is nearly
independent of the temperature and changes slowly in agreement with \cite{bayat}.\\

 \begin{figure}
\centering
    \includegraphics[width=7cm,height=5cm,angle=0]{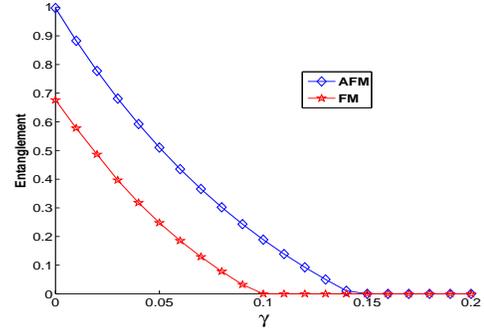}
    \caption{(Color online) The amount of entanglement in its first maximum between two ends in a chain of length 6 versus the
    decoherence parameter $\gamma$ in a fully polarized
    environment for both the case of FM chain and AFM one.}
    \label{fig4}
\end{figure}
In practical situations it is impossible to isolate the quantum
systems from their environment. In the case of Markovian interaction
between system and the environment a Lindblad equation  describes
the evolution of the system: $\dot{\rho}=-i[H,\rho]+\ell (\rho)$,
where $\ell(\rho)$ is the Markovian evolution of the state $\rho$.
In context of the situation we are studying it is reasonable to
assume an environment which has no preferred direction. It is
precisely for such an environment that a stable $SU(2)$ symmetric
AFM ground state makes sense. Otherwise, for example, in an
environment where some spin direction spontaneously decays to its
opposite direction (an amplitude damping environment, in other
words), an AFM state with approximately half the spins facing
opposite to each other will decay into a symmetry broken FM ground
state. Then the very premise of our investigation, namely starting
from a AFM ground state looses meaning. Thus it is a reasonable
assumption that the non unitary evolution $\ell(\rho)$ has the form
\begin{eqnarray}\label{Leinblad}
    \ell(\rho)=-\frac{\gamma}{3} \sum_i\sum_{\alpha}\{\rho-\sigma_{\alpha i}\rho\sigma_{\alpha
    i}\},
\end{eqnarray}
where index $i$  takes $0',0,...,N_{ch}$ and $\alpha$ gets
$x,y,z$. The operators $\sigma_{\alpha i}$ means that the operator
$\sigma_\alpha$, which can be any of Pauli matrices, acting on the
$i$th site of the whole system. The coefficient $\gamma$ stands
for the rate of decoherence in this dissipative environment. In
figure \ref{fig4}, we have plotted the first maximum of entanglement versus
 $\gamma$ for both the case of FM and AFM
 chains. In both cases, the entanglement decays
 exponentially with the decoherence parameter $\gamma$ but the FM chain decays much more faster
 than AFM chain.
\begin{figure}
\centering
    \includegraphics[width=8cm,height=7cm,angle=0]{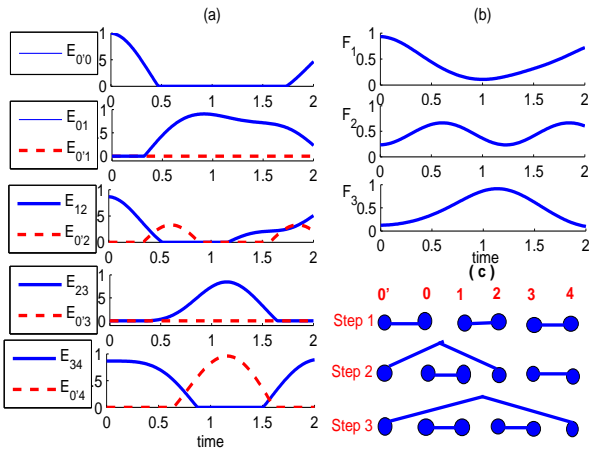}
    \caption{(Color online) The entanglement between site $0'$ and the other sites $0,1,2,...,Nch$ during the
     time evolution in an AFM chain of length $N=6$.}
     \label{fig5}
\end{figure}

  One can spot a simple but curious physical picture
which describes the propagation of entanglement through the chains with even
number of spins. Firstly, note that though one has the simplest
possible spin-1/2 AFM chain (a uniformly coupled nearest neighbor
chain) where one does not normally expect a dimer phase, the ground
state is somewhat dimerized because of the ``open ends" \cite{wang}.
Thus if one takes an approach whereby one draws a bond for the
presence of strong entanglement and no bond for very weak entanglement ($<0.1$), the
open ended AFM chain will be depicted as a dimerized state (though
it is far from being an exact dimer). Appending a singlet of spins
$0$ and $0'$ at one end of the chain, makes the total system look
like a series of strongly entangled pairs next to each other and
this is shown for the $N=6$ case in step 1 of Fig. \ref{fig5} c. The
entanglements between $0'$ and any of the other spins of the chain,
as well as the entanglement existing between the nearest neighbors
for this chain are plotted in Fig. \ref{fig5}a. Surprisingly there
is no entanglement at any time between site $0'$ and odd sites of the chain.
The mode of propagation of entanglement through the spin chain is thus
depicted in steps 1-3 of Fig. \ref{fig5} c. Note that a bond drawn
between site $0'$ and any of the other spins shown in the figure
truly corresponds to the presence of entanglement between $0'$ and that spin
(in other words, it is absent if there is no bond). Step 1 is an
approximation of the initial state, while steps 2 and 3 are the
times that spin $0'$ gets entangled with spins 2 and 4 respectively.
The dynamics of entanglement of $0'$ hopping along the chain to skip alternate
sites is generic for all even chains that we have considered. For
the simplest case of $N=4$, which can be analytically computed, the
entanglement dynamics is simply an sinusoidal oscillation between the two
states $|\psi^{-}\ra_{0'0}|\psi^{-}\ra_{12}$ and
$|\psi^{-}\ra_{0'2}|\psi^{-}\ra_{01}$ with frequency $2Jt$ (a
similar effect has been seen for spin-1 dimers and trimers in
Ref.\cite{sanpera}). It is a generalization of this effect that we
see for higher $N$. The curious dynamics depicted in figure
\ref{fig5}c is, in fact, a very good approximation of the true
dynamics even if the bonds were thought of as real singlets, and the
overlap of that approximation with the ``true" dynamics is shown in
figure \ref{fig5}b.

  In this paper we have examined the
transfer of entanglement through AFM spin chains and found peculiar features
including a nonanalytic behavior in the time variation of the
transferred entanglement and a curious hopping mode of entanglement propagation skipping
alternate sites of the channel. These predictions should be very
interesting to test, potentially through local measurements on spins
that can witness entanglement in an experiment (one such example requires
classically correlated measurements of spin operators in only three
directions \cite{guhne}), especially through NMR experiments
\cite{NMR,AFM-Polymer} or fabricated AFM nano-chains \cite{hirjib}.
The amount of entanglement, purity and also its velocity of distribution in
AFM is found to be superior to the case of FM chains, as well as the
states being readily distillable. Furthermore AFM chains are more
resistive to temperature and decoherence effects. It is an open
question whether any of the plethora of techniques for perfecting
the entanglement transfer in FM chains, such as coding and engineering
\cite{rest,rest1}, have AFM analogs.

SB is supported by an Advanced Research Fellowship from EPSRC,
through which a part of the stay of AB at UCL is funded, and the
QIP IRC (GR/S82176/01). AB thanks the British Council in Iran for
their scholarship.

\end{document}